\documentstyle{aipproc}  %[12pt]{article}
\begin{document}
\title{Aspects of Charmonium Physics}
\author{S.F.~Tuan}
\address{Department of Physics, University of Hawaii at Manoa \\ 
Honolulu, HI 96822-2219, U.S.A.}
\maketitle
\begin{abstract}
I review possible resolution of the $J/\psi (\psi^{\prime}) \rightarrow
\rho-\pi$ puzzle based on two inputs: the relative phase between the one-
photon and the gluonic decay amplitudes, and a possible hadronic excess
in the inclusive nonelectromagnetic decay rate of $\psi^{\prime}$. The 
status of a universal large phase here is examined for its meaning and
implications (including those for B-physics). Since the future of 
tau/charm facility(s) are again under consideration together with a future
anti-proton facility at GSI, I propose to extend my review to include a
broader discussion of charmonium physics. Outstanding questions like the
status of the $^{1}P_{1}$ state of charmonium, measuring $D^{0}-\bar D^{0}$
mixing and relative strong phases, status of molecular P(S) - wave charmonia
will also be discussed amongst others.
\end{abstract}

The status of the famous $J/\psi(\psi^{\prime}) \rightarrow \rho-\pi$ puzzle
has been summarized well\cite{REF1} elsewhere, including the pertinent 
experimental features. By the time of the Hadron'99 Conference in Beijing it
was known\cite{REF2} that almost all theoretical models aiming to explain
the puzzle, were found to be inadequate. Gu and Li\cite{REF3} stressed that
a key premise for physical considerations is to establish whether the
$J/\psi$ decays or the $\psi(2S)/\psi^{\prime}$ decays are anomalous. This is
particularly relevant relative to a more recent model\cite{REF4} where it
was proposed that the $J/\psi \rightarrow \rho\pi$ decay is {\em enhanced}
(rather than $\psi^{\prime} \rightarrow \rho\pi$ being suppressed) by mixing
of the $J/\psi$ with light-quark states, notably $\omega$ and $\phi$. 
Arguments against the model are given by Rosner \cite{REF1} 
include the appearance of certain
unsuppressed light-quark decay modes of the $\psi^{\prime}$ and the lack of
evidence for helicity suppression in $J/\psi$ decays involving a single
virtual photon. Indeed an amplitude analysis made for the two-body decays
of $J/\psi$ to VP\cite{REF5} has shown that nothing anomalous is found in
the magnitudes of the three-gluon and one-photon amplitudes. Hence those 
arguments presupposing the $J/\psi$ as the origin of the puzzle anomaly
should be disregarded, including another model presented at
Hadron'99\cite{REF6}.

In order to search for a clue to solve the $\rho\pi$ puzzle, Suzuki\cite{REF7}
argued with conviction that we must locate the source of the problem before
offering a final solution of the problem. In other words we must first identify
the correct inputs towards resolution. Two threads have been exposed clearly
since Hadron'99 which may eventually lead to a solution of the $\rho\pi$
puzzle. Motivated by the results in the amplitude analyses\cite{REF5,REF7} of
the two-body decays of $J/\psi$, the following two input postulates seem
eminently reasonable. (1) The relative phases between the gluon and the 
photon decay amplitudes are {\em universally} large (close to $\pi/2$) for 
all two body decays of $J/\psi$. The photon decay amplitudes are 
predominantly real and consequently the gluon decay amplitudes are imaginary.
The same pattern holds for $\psi(2S)$ decay as well. (2) A possible hadronic
excess in the inclusive nonelectromagnetic decay rate of $\psi(2S)$.

The large phase assumption (1) is based on the amplitude analysis of the
$J/\psi$ decay where the relative phase of the gluonic and the one-photon 
decay amplitude is close to $90^{o}$ for all two-body decay channels so far
studied\cite{REF7}: $1^{-}0^{-}$, $0^{-}0^{-}$, $1^{-}1^{-}$, $1^{+}0^{-}$,
and $N\bar{N}$. The appropriate references are given in ref. [7], to which we
add the almost model independent work of Achasov and Gubin\cite{REF8} on
large (nearly $90^{o}$) relative phase between the one-photon and the 
three-gluon decay amplitudes in $J/\psi \rightarrow \rho\eta$ and 
$\omega\eta$ decays. Of particular interest is that the nucleon-antinucleon
FSI large phase is based purely on {\em experiment} of the FENICE 
Collaboration. As an historical note it was pointed out to me by Achasov\cite{REF9}
that more than 10 years ago MARK III\cite{REF10} and DM2\cite{REF11} teams
as well as numerous theoretical users missed a top level result by failing
to decode their data on the strong evidence (if not discovery) of a large
(near $90^{o}$) phase between the $J/\psi \rightarrow \rho\eta$ (one-photon)
and the $\omega\eta$ (three-gluon) amplitudes! Assumption (2) is based on
experimental information relevant to the issue. That is the hadronic decay 
rate of $\psi(2S)$ which is normally attributed to $\psi(2S) \rightarrow ggg$.
When we compute the current data for the inclusive gluonic decay rate of
$\psi(2S)$ by subtracting the cascade and the electromagnetic decay rate from
the total rate, it is 60-70\% larger, within experimental uncertainties, than
what we expect from short-distance (i.e. computable from perturbative QCD)
gluonic decay alone - the celebrated 12-14\% rule. Smaller errors have been
attached to this discrepancy\cite{REF3} with a different error estimate, but
the conclusion remains basically the same. 

On the matter of `universal large phases', this was argued recently by 
G\'{e}rard and Weyers\cite{REF12} in their model. Some of the experimental
and theoretical difficulties for $\rho-\pi$ puzzle resolution via this model
have already been pointed out by Gu and Tuan\cite{REF2}. While agreeing that
there could be a more universal nature to the conclusion that the photon and
three-gluon amplitudes for $J/\psi$ are out of phase with one another,
Rosner\cite{REF13} found their argument that the three-gluon and photon final
states are orthogonal somewhat curious. They can populate the same hadronic
final states (with the same spin and parity). Nothing then in principle 
would prevent a real relative phase between a three-gluon and a one-photon 
amplitude. Thus the argument (based on incoherence) for this large relative
phase leaves something to be desired. Nevertheless, without over emphasizing
the large universal phase, Rosner\cite{REF14} pointed out that in the charm
decays themselves, even the $\rho-K$ final state had large relative phases
among different amplitudes, despite the fact that these did not show up in 
the isospin triangle discussed earlier\cite{REF15} as an aftermath of Suzuki's
analysis\cite{REF5}.

In point of fact, the decay branching fractions of $\psi(2S) \rightarrow
1^{-}0^{-}$ clearly show a suppression of the gluon amplitude and favor a
{\em small} relative phase between the gluon and photon amplitudes\cite{REF7}.
A small phase seems likely for $\psi(2S) \rightarrow 1^{+}0^{-}$ also.
At first sight this seems a violation of assumption (1) above. However taking
into account assumption (2) about the possible excess in the inclusive 
hadronic decay rate of $\psi(2S)$, it is proposed that this excess is related
to both the suppression and the {\em small} relative phase of the $1^{-}0^{-}$
amplitude which would otherwise have the {\em large} relative phase of 
assumption (1). The proposition is then that an additional decay process 
generating the excess should largely cancel the short-distance gluon amplitude in
the exclusive decay into $1^{-}0^{-}$ and that the resulting small residual
amplitude is not only {\em real} but also destructively interferes with the
photon amplitude. Remember\cite{REF16} the picture is still that the three-
gluon decay amplitudes of $J/\psi$ and $\psi^{\prime}$ have large phases 
because the three gluons are on mass shell. In contrast, one-photon 
annihilation amplitudes are real because the photon is off shell. Suzuki\cite{REF7}
then examined two scenarios which may possibly generate the excess inclusive
hadronic decay. They seem to be among a very few possibilities that
have not yet been ruled out by experiment. We summarize these model scenarios
in the sections below. 

The $\psi(2S) \rightarrow resonance \rightarrow hadrons$ is a twist of an old 
one: A noncharm resonance may exist near the $\psi(2S)$ mass and give an 
extra contribution to the hadronic decay rate. A glueball was proposed earlier
at the $J/\psi$ mass to boost the $\rho\pi$ decay of $J/\psi$\cite{REF17}.
However we now want it near $\psi(2S)$, not near $J/\psi$. Suzuki\cite{REF7}
argued against this resonance, call it R, being a $q\bar{q}$ or four-quark
resonance. For a $1^{--}$ glueball, the lattice calculations\cite{REF18}
indeed support such a state at 3.7 GeV or even higher. However\cite{REF16} it
is always nice to get a physical picture from some analytic calculation.
Lattice calculation may be fine but only after one gets a reasonable 
quantitative understanding without it. It is reassuring that a very recent work by
Hou {\em et al.}\cite{REF19} using the constituent gluon model estimate the
mass of $1^{--}$ glueball to be 3.1 - 3.7 GeV, which is close to the mass of
the $J/\psi$ and $\psi^{\prime}$. We look into the possibility that R around
the $\psi(2S)$ mass destructively interferes with the perturbative $\psi(2S)
\rightarrow ggg \rightarrow 1^{-}0^{-}$ decay. Note unlike the glueball 
proposed near $J/\psi$ mass to enhance $J/\psi \rightarrow \rho\pi$\cite{REF17}
and hence the unnatural property that it decays predominantly into 
$1^{-}0^{-}$, in our case, since glueball R is introduced to account for the
hadronic excess, it should couple not primarily to the $1^{-}0^{-}$ channels,
but to many other channels and hence is quite natural. If R has total width
$\Gamma_{R}$ is as narrow as 100 MeV, for instance, the mixing $|\epsilon|$
= {\em O}$(10^{-2})$ would be able to account for the excess in the inclusive
hadron decay of $\psi(2S)$. It was also argued\cite{REF7} that the amplitude
for $\psi(2S) \rightarrow R \rightarrow$ hadrons has automatically a large
phase when mass difference between R and $\psi(2S)$ is smaller than
$\Gamma_{R}$. Hence the resonant amplitude can interfere strongly with the
three-gluon amplitude in two-meson decays. However Hou\cite{REF20} has argued
that the glueball could be significantly narrower than 100 MeV using the
``[OZI]$^{1/2}$'' rule\cite{REF21} that a typical glueball R total width is
$\Gamma_{R} = [\Gamma_{\rm ordinary} \times \Gamma_{\rm OZI-violating}]^{1/2}$. In
terms of $\psi(2S)$ where $\Gamma_{ordinary}$ could be generously up to 500
MeV in width at this mass, while $\Gamma(\psi(2S))$ is about 277 KeV, the
square root formula gives for a gluonium R degenerate with $\psi(2S)$ a width
of order 11.8 MeV. Using Meshkov's\cite{REF21} benchmark, something like 35.4
MeV would be a conservative estimate. It would be interesting for BES to
proceed from their 3.96 million $\psi(2S)$ to search for indirect evidence 
of R via scanning across $\psi(2S)$ for shape distortion, but even the
setting of a limit on the gluonium width will be of interest. We need of
course to work with unsuppressed $\psi(2S) \rightarrow PA$ modes 
($\pi b_1(1235)$, $K_1(1270)\bar{K})$ or even the unsuppressed SV mode
$f_{0}\phi$ of $\psi(2S)$. If the glueball is significantly narrower than
100 MeV, the mixing parameter $\epsilon$ must be adjusted accordingly. The
adjustment depends on how closely $\psi(2S)$ and the glueball R are 
degenerate. How much this affects the $\psi(2S)$ resonance shape depends
sensitively on how closely $\psi(2S)$ and the glueball are degenerate. To
observe a distortion of the resonance shape of $\psi(2S)$, they would have to
be degenerate nearly to the $\psi(2S)$ width, 277 KeV!

The $\psi(2S) \rightarrow D\bar{D} \rightarrow$ hadrons model\cite{REF7} has 
been much expanded by Rosner\cite{REF22} very recently. He noted that if a
$\psi^{\prime}$ decay amplitude due to coupling to virtual (but nearly on
shell) charmed particle pairs interferes destructively with the standard
three-gluon amplitude, the suppression of these (and possibly other) modes
in $\psi^{\prime}$ final states can be understood. However Rosner goes 
further and noted effects of the proximity of the $D\bar{D}$ threshold can 
mix the $\psi^{\prime}$ and the $\psi(3770)$. Perhaps the missing partial
width of $\psi^{\prime} \rightarrow \rho\pi$ (less than half a KeV) is showing
up in the $\psi(3770)$. If so, since the latter state has a total width
nearly 100 times that of the $\psi^{\prime}$, it would correspond to a very
tiny branching ratio yet to be detected. As pointed out\cite{REF7} the
$D\bar{D}$ amplitude needs to have a large final-state phase, in order to
interfere destructively with the perturbative 3g contribution in the
$\rho\pi$ and $K\bar{K}^{*}(892)$ + c.c. channels. If this new contribution is
due to rescattering into non-charmed final states through charmed particle
pairs, it is exactly the type of contribution proposed by many and quoted in
Ref. [22], in which the decay $\bar{b} \rightarrow \bar{c}c\bar{s}$ or $\bar{b}
\rightarrow \bar{c}c\bar{d}$ contributes to the penguin amplitude {\em with a
large phase}. The rescattering of $D\bar{D}$ states into non-charmed final
states (like $\rho-\pi$) could also be responsible for the larger-than-
expected penguin amplitude in B decays about which the Rome people\cite{REF23}
have been writing about, and for the large $B \rightarrow K\eta^{\prime}$
branching ratio. If the phase is large, as needed for a suppression of 
$\psi^{\prime} \rightarrow \rho\pi$, this again would be good news for the
observation of a large CP asymmetry in B decays with both tree and penguin
contributions\cite{REF15}. Hence there is cautious optimism that the 
amplitude for $\psi(2S) \rightarrow D\bar{D} \rightarrow mesons$ can have the
requisite large size and phase to help resolve the $\rho-\pi$ puzzle.

{\bf Remarks:} (a) It must be recognized that both models above, resonance
R or $D\bar{D}$ scattering, involves long distance effects not easily 
computable, unlike the short distance perturbative QCD. (b) The situation for
$\psi(2S) \rightarrow VT$ (suppressed), $\psi(2S) \rightarrow PA$ 
(unsuppressed for $\pi b_{1}(1235)$, $K_{1}(1270)\bar{K})$, and $\psi(2S)
\rightarrow SV$ (unsuppressed for $f_{0}\phi)$ exclusive channels deserve
further study as the first step (we recognize other anomalies,
e.g. anomalous enhancement of $\psi (2S) \rightarrow K_1 (1270)
\bar{K}$, different isospin violations in $K^* \bar{K}$ decays for
$J/\psi$ and $\psi (2S)$, flavor-SU(3)-violating $K_1 (1270) -K_1
(1400)$ asymmetries with opposite character for 
$J/\psi$ and $\psi(2S)$, which need to be taken up later). At present we have no way to 
relate among two-body charmonium
decay amplitudes of different spin-parities. Probably we shall not have one
for a long time. This is a weak point of the argument of ``long-distance
physics'' in contrast to the short-distance argument such as ``helicity
suppression'' of perturbative QCD. It would be good to have some systematic
``long-distance'' argument which covers two-body decays of different 
spin-parities in one shot! (c) The decay angular distributions for $1^{-}0^{-}$
and $0^{-}0^{-}$ should be tested. A large interference can occur only when
the dynamical mechanisms of the two processes are similar. When a large
disparity is observed between the corresponding two-meson decay rates of
$J/\psi$ and $\psi(2S)$, the decay angular distribution of this channel will
also be very different between $J/\psi$ and $\psi(2S)$. This will give a
good test of the idea of interference with an additional amplitude. However
in the PP and VP decays, the decay angular distribution is unique, 
independent of dynamics, because there is only one relative orbital angular momentum
involved. That is not the case for the decays such as VT: the final orbital
angular momentum can take different values (0, 2, or 4 for VT). This actually
opens up the possibility to test Suzuki's\cite{REF7} dynamical assumptions
by concentrating on PP decays especially. As pointed out \cite{REF3}
since $\psi (2S) \rightarrow PV$ strong decay is much suppressed,
measurement of its decay angular distribution would be extremely
difficult.  However $\psi (2S) \rightarrow PP$ is already giving
tantalizing hint of constructive interference in (large) decay rates
\cite{REF7} and should be checked in decay angular distribution. 
(d) Unlike for the charmonium case
where hadron helicity conservation (HHC) and perturbative QCD (PQCD) is of
doubtful validity\cite{REF24}, HHC/PQCD at the $\Upsilon(nS)$ mass are more
immune from breakdown. Mechanisms for possible breakdown at $\psi(nS)$
energies are minimized at the heavier $b\bar{b}$ mass. For $\Upsilon(nS)$
exclusive decays to light hadrons, violations are of order $m_{h}/m_{Q}$
(light hadron mass/heavy quark mass) in  matrix element; 
or of order $(m_{h}/m_{b})^{2}$ if the
light hadron has intrinsic $b\bar{b}$ content in decays\cite{REF16}. The
$c \bar{c}$ contribution from light hadrons in upsilon exclusive decays
should also be not very important, such as for $\Upsilon(nS) \rightarrow
\rho \pi$, since the probability of the intrinsic bottom in the $\rho$
and $\pi$ is suppressed by $(m_{c}/m_{b})^{2}$ relative to the intrinsic
charm probability.  This is shown rigorously using the operator product
expansion OPE\cite{REF25}.  Both these 
corrections are negligible for $\Upsilon(nS)$. Nevertheless it would be of
interest to test $\Upsilon(nS) \rightarrow \omega-\pi^{0}$ (I=1 
electromagnetic) decay strength, to reassure us of the absence of long distance
effects at $\Upsilon(nS)$. (e) At a deeper level Kochelev\cite{REF26} has 
advanced the instanton approach (shared by M.A. Shifman) for understanding
the $\rho-\pi$ puzzle. He noted that if one consider the decay 
$J/\psi(\psi(2S)) \rightarrow ggg$ then the average virtuality of each gluon
is approximately $(2m_{c}/3)^{2} \approx 1 GeV^{2}$. It is difficult to
believe that at such virtuality one can apply perturbative QCD. He has
estimated the direct instanton contribution to nucleon sum rules\cite{REF27}
where he showed that at scale $M_{P}^{2} \approx 1 GeV^{2}$ the instanton
contribution is very important. But the virtuality of each gluon for the
$\Upsilon(nS)$ will be even larger. An interesting question is whether the
relative phases are close to $90^{o}$, with photon decay amplitudes real and
consequently the gluon decay amplitudes are imaginary for $\Upsilon(nS)$ as
once conjectured by Suzuki\cite{REF16}.

Since the future of tau/charm facility(s) are again under consideration
together with a future anti-proton facility at GSI, I shall very briefly list
some broader aspects of charm/charmonium physics which can be accomplished
when such facilities are available. (i) The study of $\psi(^{1}P_{1})$ state
should be vigorously pursued. Rosner's mixing solutions\cite{REF22} are in
tantalizing agreement with those of Kuang-Yan\cite{REF28} and make the case
for charmonium $^{1}P_{1}$ even more exciting. Other details are given
elsewhere\cite{REF29}. (ii) The measuring of $D^{0}-\bar{D}^{0}$ mixing and
relative strong phases at a Charm Factory has been discussed recently by
Gronau {\em et al.}\cite{REF30}. (iii) The Martenelli lattice group in Rome
(with their more powerful computing facility) should be encouraged to reach
a conclusion whether the Alford/Jaffe\cite{REF31} prediction of a $J^{PC}$ =
$0^{++}$ stable S-wave four-quark bound state, with non-exotic flavor quantum
numbers, just below the threshold for $D\bar{D}$ in the charmonium spectrum,
is sustainable or not. Though P-wave molecular charmonium states\cite{REF32},
because of centrifugal barrier and hence less quark wavefunction overlap for
formation\cite{REF33}, should also be tested by the lattice group for 
sustainability. 

I wish to thank my scientific colleagues Kolia Achasov, Jon Rosner, and
Mahiko Suzuki for very helpful communications and discussions. This work was
supported in part by the U.S. Department of Energy under Grant DE-FG-03-
94ER40833 at the University of Hawaii at Manoa.

%\end{section}


\begin{thebibliography}{99}
\bibitem{REF1} See for instance F.A.~Harris, hep-ex/9903036, March 1999;
J.Z. Bai {\em et al.}, {\em Phys. Rev. Lett.} {\bf 83}, 1918 (1999);
Y.F. Gu, {\em Proc. of DPF '96}, {\bf Vol. 2}, p. 986 [World Scientific Publishing
(1998)]; J.L. Rosner, see Ref. 22 below.
\bibitem{REF2} Y.F.~Gu and S.F.~Tuan, {\em Nucl. Phys.} {\bf A675}, 404 (2000); 
S.F.~Tuan, {\em Commun. Theor. Phys.} {\bf 33}, 285 (2000).
\bibitem{REF3} Y.F.~Gu and X.H.~Li, {\em Phys. Rev.} {\bf D63}, 114019 (2001).
\bibitem{REF4} T.~Feldmann and P.~Kroll, {\em Phys. Rev.} {\bf D62}, 074006 (2000).
\bibitem{REF5} M.~Suzuki, {\em Phys. Rev.} {\bf D57}, 5717 (1998).
\bibitem{REF6} C.T.~Chan and W.S.~Hou, {\em Nucl. Phys.} {\bf A675}, 367(2000).
\bibitem{REF7} M.~Suzuki, {\em Phys. Rev.} {\bf D63}, 054021 (2001). 
\bibitem{REF8} N.N.~Achasov and V.V.~Gubin, {\em Phys. Rev.} {\bf D61}, 117504 (2000).
\bibitem{REF9} N.N.~Achasov, private communication.
\bibitem{REF10} MARK III Collaboration, {\em Phys. Rev.} {\bf D38}, 2695 (1988).
\bibitem{REF11} DM2 Collaboration, {\em Phys. Rev.} {\bf D41}, 1389 (1990).
\bibitem{REF12} J.-M.~G\'{e}rard and J.~Weyers, {\em Phys. Lett.} {\bf
B462}, 324 (1999).
\bibitem{REF13} J.L.~Rosner, private communication.
\bibitem{REF14} J.L.~Rosner, {\em Phys. Rev.} {\bf D60}, 114026 (1999).
\bibitem{REF15} J.L.~Rosner, {\em Phys. Rev.} {\bf D60}, 074029 (1999).
\bibitem{REF16} M.~Suzuki, private communication.
\bibitem{REF17} S.J.~Brodsky {\em et al.}, {\em Phys. Rev. Lett.} {\bf 59},
 621 (1987).
\bibitem{REF18} M.~Peardon, hep-lat/9710029; C. Morningstar and
M. Peardon,hep-lat/9901004 v.2; C. Morningstar, {\em these Proceedings}.
\bibitem{REF19} W.S.~Hou {\em et al.}, {\em Phys. Rev.} {\bf D64},
014028 (2001).
\bibitem{REF20} W.S.~Hou, private communication.
\bibitem{REF21} D.~Robson, {\em Nucl. Phys.} {\bf B130}, 328 (1977); S.~Meshkov
{\em et al.}, {\em Phys. Lett.} {\bf 99B}, 353 (1981); S.~Meshkov, CALT-68-923,
UCI Technical Report 82/35 (1982).
\bibitem{REF22} J.L.~Rosner, hep-ph/0105327.
\bibitem{REF23} M.~Ciuchini {\em et al.}, {\em Nucl. Phys.} {\bf B501},
271 (1997); M.~Ciuchini, R.~Contino {\em et al.}, {\em Nucl. Phys.} {\bf
B512}, 3 (1998).
\bibitem{REF24} V.~Chernyak, hep-ph/9906387.
\bibitem{REF25} S.J.~Brodsky, private communication concerning the work
of M.~Franz, M.V. Polyakov, and K. Goeke, {\em Phys. Rev.} {\bf D62},
074024 (2000).
\bibitem{REF26} N.~Kochelev, private communication.
\bibitem{REF27} N.~Kochelev, {\em Zeit. Phys.} {\bf C46}, 281 (1990).
\bibitem{REF28} Y.P.~Kuang and T.M.~Yan, {\em Phys. Rev.} {\bf D41}, 155 (1990).
\bibitem{REF29} T.~Barnes {\em et al.}, UH-511-868-97, Ms. \# 8921 at PLB.
\bibitem{REF30} M.~Gronau {\em et al.}, hep-ph/0103110.
\bibitem{REF31} M.~Alford and R.L.~Jaffe, hep-lat/0001023, MIT-CTP-2940.
\bibitem{REF32} A.~De R\'{u}jula {\em et al.}, {\em Phys. Rev. Lett.} {\bf 38},
317 (1977); S.F.~Tuan, {\em Phys. Lett.} {\bf B473}, 136 (2000).
\bibitem{REF33} A.~De R\'{u}jula and R.L.~Jaffe, {\em Experimental Meson
Spectroscopy} (1977), p. 83 [Published by Northeastern University Press, Boston,
Mass 02115].

\end{thebibliography}
\end{document}